\documentclass[12pt]{article}

\usepackage{amsmath,amssymb}
\usepackage{cite}

\usepackage[unicode,linktocpage=true,plainpages=false,pdfpagelabels=false]{hyperref}

\newcommand{\dd}{\partial}
\newcommand{\de}{\delta}
\newcommand{\m}{\mu}
\newcommand{\n}{\nu}
\newcommand{\ls}{\left(}
\newcommand{\rs}{\right)}
\newcommand{\la}{\lambda}
\newcommand{\ka}{\varkappa}
\newcommand{\ga}{\gamma}
\newcommand{\si}{\sigma}
\newcommand{\ta}{\tau}
\newcommand{\al}{\alpha}
\newcommand{\be}{\beta}

\newcommand{\disn}[2]{$$\displaylines{\refstepcounter{equation}%
            \label{#1}\hskip 1em minus 1em #2\hfilneg}$$}
\newcommand{\nom}{\hfil\hskip 1em minus 1em (\theequation)}

\newcommand{\ns}{\hfill\cr\hfill}

\begin{document}

\title{Embedding theory as new geometrical mimetic gravity}
\author{
S.~A.~Paston\thanks{E-mail: s.paston@spbu.ru},
A.~A.~Sheykin\thanks{E-mail: a.sheykin@spbu.ru}\\
{\it Saint Petersburg State University, Saint Petersburg, Russia}
}
\date{\vskip 15mm}
\maketitle

\begin{abstract}
It is known that recently proposed model of mimetic gravity can be presented as general relativity with an additional mimetic matter.
We discuss a possibility to analogously reformulate the embedding theory, which is
the geometrical description of gravity proposed by Regge and Teitelboim, treating it also as general relativity  with some additional matter.
We propose a form of action which allows to describe this matter in terms of conserved currents. This action turns out to be a generalization of the perfect fluid action, which can be useful in the analysis of the properties of the additional matter. On the other side, the action contains a trace of the root of the matrix product, which is similar to the constructions appearing in bimetric theories of gravity. The action is completely equivalent to the original embedding theory, so it is not just some artificial model, but has a clear geometric sense. We discuss the possible equivalent forms of the theory and ways of study of the appearing equations of motion.
\end{abstract}

\newpage

\section{Introduction}\label{emmg-vved}
The idea of \emph{mimetic gravity} was proposed five years ago in the paper \cite{mukhanov} and became quite popular in recent years.
In this framework the conformal degree of freedom of gravity is isolated
by introducing a parametrization of the physical metric in the following form:
\disn{v1}{
g_{\m\n}=\tilde g_{\m\n}\tilde g^{\ga\de}(\dd_\ga\la)(\dd_\de\la).
\nom}
The authors use the familiar form of the General Relativity (GR) action which consists of the Einstein-Hilbert gravitational action \disn{vi15}{
S^{\text{EH}}[g_{\m\n}]=-\frac{1}{2\ka}\int\! d^4x\sqrt{-g}\,R(g_{\m\n})
\nom}
with respect to physical metric $g_{\m\n}$, and the matter action $S_{\text{m}}[g]$. However, independent variables in this action are auxiliary metric $\tilde g_{\m\n}$ and the scalar field $\la$ rather than $g_{\m\n}$.
The field equations are
\disn{vi18}{
G^{\m\n}=\ka \ls T^{\m\n}+n u^\m u^\n\rs,\qquad
D_\m (n u^\m)=0,\qquad
g^{\m\n}u_\m u_\n=1
\nom}
($T^{\m\n}$ is the energy momentum tensor (EMT) of matter, $D_\m$ is the covariant derivative, signature is $+---$)
where the following notation is used:
\disn{vi17}{
n\equiv g_{\m\n}\ls \frac{1}{\ka}G^{\m\n}-T^{\m\n}\rs,\qquad
u_\m\equiv\dd_\m\la.
\nom}
Since the field equations of mimetic gravity have the form of Einstein equations with an additional contribution to EMT corresponding to \emph{mimetic matter}, one can say that  "{}`dark matter` without dark matter, which is imitated by extra scalar degree of freedom
of the gravitational field"{} \cite{mukhanov} arises in the theory.

As can be seen from the field equations, the appearing mimetic matter has the form of \textit{pressure-free perfect fluid} with \textit{potential} motion. It allows one to write an equivalent formulation of mimetic gravity, where physical metric is considered to be independent, and an additional action term corresponding to such perfect fluid is present. For the first time it was done in the paper
\cite{Golovnev201439}, where the full action is chosen to be
\disn{v2}{
S=S^{\text{EH}}+S_{\text{m}}+S^{\text{add}},
\nom}
where
\disn{p12}{
S^{\text{add}}=-\frac{1}{2}\int\! d^4 x\, \sqrt{-g}\,\Bigl( 1-g^{\m\n}(\dd_\m\la)(\dd_\n\la)\Bigr)n.
\nom}
In such framework the mimetic matter is described by two scalar fields: $n$ can be treated as number density
of mimetic matter particles and $\la$ defines their velocity $u_\m=\dd_\m\la$.
There are other ways of choosing the additional term in action for potentially moving pressure-free perfect fluid, see \cite{statja48}.
There are also many other ways to write an action for mimetic gravity,
see, e.g., \cite{arXiv1311.3111,arXiv1512.09118}.

The framework of mimetic gravity can be modified to explain certain phenomena of modern cosmology. In particular, the introduction  of a potential for the scalar field $\la$ \cite{mukhanov2014} leads to the appearing of a pressure of the mimetic matter
(note that used structure of the action appears to be a particular case of the
general expression which was already considered in the earlier work \cite{lim1003.5751}),
whereas an addition of a higher-order derivative term transforms mimetic matter into imperfect fluid \cite{mukhanov2014,vikman2015,kobayashi2017}. It is also possible to write an action of mimetic matter in a form with which it turns out to be pressure-free perfect fluid moving arbitrary (i.e. not potentially anymore) \cite{statja48}. In the latter case the description of mimetic matter contains two more fields in addition to $n$ and $\la$. Moreover, one can return to the original formulation of mimetic gravity with the auxiliary  metric $\tilde g_{\m\n}$ and three scalar fields that are present in the expression for physical metric $g_{\m\n}$ analogous to \eqref{v1}, see details in \cite{statja48}. For the current status of mimetic gravity framework see the review  \cite{mimetic-review17} and the references therein.

The transformations \eqref{v1} on which the mimetic gravity is based are the special case of the more general "disformations"{} \cite{Bekenstein1993}, which are invertible in general case, i.e. auxiliary metric $\tilde g_{\m\n}$ can be expressed through
the physical metric $g_{\m\n}$ and scalar field $\la$. It leads to the fact that in the general case of such a change of variables the theory turns out to be equivalent to GR \cite{arXiv1407.0825}.
On the contrary, the special transformations \eqref{v1} is non-invertible because
the conformal mode  of $\tilde g_{\m\n}$ does not contribute to $g_{\m\n}$, i.e. the conformal invariance appears \cite{arXiv1311.3111}.
Hence the mimetic gravity is an example of a theory modification appearing as a result of the change of variables that contains differentiation while the number of variables remains unchanged as 10 components of $g_{\m\n}$ is replaced by  9 components $\tilde g_{\m\n}$
(without the conformal mode) and the scalar $\la$.
This modification is caused by the change of the class of variations of independent field variables, which alters the set of possible solutions of the theory according to the property of the variational principle. In the case of the change of variables that do not contain a differentiation the change of the class of variations does not occur (of course, if the change of variables does not affect the number of degrees of freedom), so the theory remains unchanged. An example of such case is a transition from the metric formulation of GR to the tetrad one. Also, if the change of variables contains only spatial derivatives, then the theory usually is not affected due to the common assumption that at the spatial infinity fields decrease sufficiently fast. Therefore the presence of time derivatives in the change of variables leads to a significant modification of the theory. Since the class of variations becomes smaller after such a change (the variation of an independent variable is always assumed to vanish at the initial and final moments of time, so this requirement put an additional restriction on the original variable), the set of solutions of modified theory turns out to be larger, including all the solutions of the original theory. It is precisely so in the case of mimetic gravity: the solutions of  \eqref{vi18} at $n=0$ (which corresponds to the absence of mimetic matter) is GR ones. A simple example of change of variables with differentiation in 0-dimensional (i.e. mechanical) theory can be found in \cite{statja33}.

Another example of change of variables in GR that contains differentiation is known for a long time: it is the Regge-Teitelboim gravity \cite{regge} also known as \emph{embedding theory}. In contrast with mimetic gravity \eqref{v1} the underlying change of variables in the embedding theory approach has a deep geometric sense. In this string-inspired approach an assumption is made that
our spacetime is not only an abstract pseudo-Riemannian space, but rather a surface in a flat $N$-dimendional ambient space (bulk). Consequently, the induced metric appears on the surface
 \disn{r1}{
g_{\m\n}=(\dd_\m y^a)(\dd_\n y^b)\,\eta_{ab},
\nom}
where $\eta_{ab}$ is a flat ambient space metric, $a,b=0,\ldots,N-1$.
This relation defines the change of variables and plays the role analogous to \eqref{v1}, expressing the physical metric through the embedding function $y^a(x)$ which turns out to be analogous to auxiliary quantities $\tilde g_{\m\n}$ and $\la$ of mimetic gravity. If one does not wish to lower the number of degrees of freedom, one must take $N\ge 10$ (since 4D metric has 10 independent components). The Friedmann-Janet-Cartan theorem \cite{fridman61} gives the same restriction: it states that an arbitrary (3+1)D pseudo-Riemannian spacetime can be locally isometrically embedded into a flat Minkowski spacetime of no less than 10 dimensions, of which at least one is timelike and at least three is spacelike. Usually (9+1)D Minkowski space is taken as ambient space of embedding theory.

Action of the embedding theory is the same expression $S[g_{\m\n}]=S^{\text{EH}}[g_{\m\n}]+S_{\text{m}}[g_{\m\n}]$ as in the mimetic gravity, but metric in it is given by \eqref{r1} instead of \eqref{v1}. Variation w.r.t. independent variable $y^a$ leads to field equations in the Regge-Teitelboim form:
 \disn{r2}{
D_\m\Bigl(( G^{\m\n}-\ka\, T^{\m\n}) \dd_\n y^a\Bigr)=0,
\nom}
which, besides GR solutions, also admit another solutions known as \textit{extra} solutions. After the original paper \cite{regge} the ideas of embedding theory were discussed in details in \cite{deser}, and later were used in the numerous papers devoted to the different aspects of gravity, including quantization, see. e.g. \cite{pavsic85let,tapia,maia89,estabrook1999,davkar,statja18,rojas09,faddeev,statja25,statja44}. A detailed bibliography of embedding theory and close topics can be found in \cite{tapiaob}.

As in mimetic gravity, the field equations \eqref{r2} of embedding theory can be written \cite{pavsic85,statja33} in the form of Eqinstein ones with an additional contribution  $\ta^{\m\n}$ to EMT which corresponds to certain fictional \emph{embedding matter}:
 \disn{r3}{
G^{\m\n}=\ka \ls T^{\m\n}+\ta^{\m\n}\rs,\qquad
\nom}
supplemented by the condition  \eqref{r1} and equation
 \disn{r4}{
D_\m\Bigl(\ta^{\m\n}\dd_\n y^a\Bigr)=0.
\nom}
Equations \eqref{r1} and \eqref{r4} can be interpreted as the equations of motion of embedding matter described by quantities $\ta^{\m\n}$ and $y^a$.
It can be shown
(see, for example, \cite{statja18})
that the standard condition of covariant conservation of embedding matter EMT $D_\m\ta^{\m\n}=0$ follows from these equations.

Several questions  can be raised: how to write an action of the embedding theory in the form of GR with additional embedding matter, what this matter is like and what are its laws of motion from the physical point of view? The most straightforward way to answer the first question was proposed in \cite{statja48}, but the choice of independent variables in there does not allow to say anything related to the second question. In the present work we will consider some alternative ways of choosing
the action of embedding theory, as well of choices of independent variables for the description of embedding matter which is more convenient in discussion of its physical meaning.

\section{Forms of action for embedding matter}
One can write such an action of embedding matter $S^{\text{add}}$, that the full action
\disn{v2a}{
S=S^{\text{EH}}+S_{\text{m}}+S^{\text{add}}
\nom}
leads to the field equations of embedding theory \eqref{r1},\eqref{r3},\eqref{r4}, in the most simple manner by satisfying the condition  \eqref{r1} through Lagrange multiplier \cite{statja48}:
\disn{za1}{
S_1^{\text{add}}=\frac{1}{2}\int\! d^4 x\, \sqrt{-g}\,
\Bigl( (\dd_\m y^a)(\dd_\n y_a) - g_{\m\n}\Bigr)\tau^{\m\n}.
\nom}
It is easy to check that the variation w.r.t. $\tau^{\m\n}$ leads to the equation \eqref{r1}, whereas w.r.t. $y^a$ -- to \eqref{r4}\footnote{Note that analogous way of addition of $S_1^{\text{add}}$ to the Nambu-Goto action leads to the Polyakov string action, see below \eqref{r19}-\eqref{r23}.
}. However, the physical sense of the variables $\ta^{\m\n}$ and $y^a$ which describe the embedding matter in such approach remains unclear.

Let us consider an alternative form of the contribution of the embedding matter to the action, namely
\disn{r5}{
S_2^{\text{add}}=\int\! d^4 x\, \sqrt{-g}\,
\Bigl( j^\m_a\dd_\m y^a-\text{\bf tr}\sqrt{g_{\m\n}j^\n_a j^{\al a}}\Bigr),
\nom}
where the operation
$\sqrt{\;\;}$ means taking a root of matrix with indices $\m$ and $\al$ and
\text{\bf tr} means subsequent trace taking.
The independent variables describing the embedding matter in this approach (we will call it \emph{embedding gravity}) are the quantities
$j^\m_a$ and $y^a$.

Let us write the appearing equations of motion. The variation w.r.t. $y^a$ immediately leads to the equation
\disn{r6}{
D_\m j^\m_a=0.
\nom}
Now let us find the variation of  \eqref{r5} w.r.t. $j^\m_a$. To do this, define
\disn{r6.01}{
A_\m{}^\al=g_{\m\n}j^\n_a j^{\al a}-\de_\m^\al.
\nom}
Also denote the result of root taking in \eqref{r5} as $\be_\m{}^\al$, so in the index-free notation
\disn{r6.02}{
\be=\sqrt{I+A},
\nom}
where $I$ is an identity matrix.
Then for the variation we have
\disn{r6.1}{
\be^2=I+A\quad\Rightarrow\quad
(\de\be)\be+\be\de\be=\de A\quad\Rightarrow\ns\Rightarrow\quad
\be^{-1}(\de\be)\be+\de\be=\be^{-1}\de A\quad\Rightarrow\quad
\de\text{\bf tr}\be=\frac{1}{2}\text{\bf tr}\ls\be^{-1}\de A\rs.
\nom}
Note that accordingly to the chosen notations the quantity $A^{\n\al}=g^{\n\m}A_\m{}^\al=j^\n_a j^{\al a}-g^{\n\al}$ turns out to be symmetric. If the matrix root is treated as a Taylor series of $\sqrt{I+A}$ w.r.t. $A$ (in assumption that root branch is taken as so that $\sqrt{I}=I$), one can easily notice that symmetricity of matrix $(g^{-1}A)$ together with symmetricity of $g$ leads to symmetricity of $(g^{-1}\sqrt{I+A})$ as all the terms in series in this case are symmetric. As a result we get $\be^{\n\al}=\be^{\al\n}$,
so the inverse of it (which we will denote as $\hat\be_{\m\n}$: $\hat\be_{\m\n}\be^{\n\al}=\de_\n^\al$) will also be symmetric: $\hat\be_{\m\n}=\hat\be_{\n\m}$. Using this fact together with \eqref{r6.1}, we obtain the equation of motion as the result of variation of \eqref{r5} w.t.t. $j^\m_a$:
\disn{r7}{
\dd_\m y^a=\hat\be_{\m\n}j^{\n a}.
\nom}
It is easy to check that the condition of the metric induceness \eqref{r1}  follows from it straightforwardly:
\disn{r8}{
(\dd_\m y^a)(\dd_\n y_a)=\hat\be_{\m\al}j^{\al a}  \hat\be_{\n\be}j^{\be}{}_a=
\hat\be_\m{}^\al(\de_\al^\be+A_\al{}^\be)\hat\be_\be{}^\ga g_{\ga\n}=g_{\m\n}.
\nom}

Now let us find the expression for EMT of embedding matter by variation of \eqref{r5} w.r.t. the metric $g_{\m\n}$. Noticing that satisfaction of EoM \eqref{r7} implies that
\disn{r9}{
j^\m_a\dd_\m y^a=j^\m_a\hat\be_{\m\n}j^{\n a}=\hat\be_\m{}^\n(\de_\n^\m+A_\n{}^\m)=
\text{\bf tr}\ls\be^{-1}(I+A)\rs=\text{\bf tr}\be
\nom}
and using \eqref{r6.1}, we obtain
\disn{r10}{
\ta^{\m\n}=\hat\be_\al{}^\m j^\n_a j^{\al a}=\ls g^{\n\al}+A^{\n\al}\rs\hat\be_\al{}^\m=\be^{\m\n}.
\nom}
Taking into account the satisfaction of the relation $\ta^{\m\n}\dd_\n y^a=j^{\m a}$ according to \eqref{r7} one can see that equation \eqref{r6} turns out to coincide with \eqref{r4}. As a result we conclude that the embedding gravity given by the action \eqref{v2a},\eqref{r5} completely reproduces the equations of motion of the embedding theory   \eqref{r1},\eqref{r3},\eqref{r4}.

The physical meaning of  the independent variables $j^\m_a$ and $y^a$ in the action of embedding matter \eqref{r5} can be understood more easily than in the action \eqref{za1}: $j^\m_a$ at the given $a$ can be treated as a conserving (according to \eqref{r6}) current density of a certain type of matter, whereas $y^a$ turn out to be Lagrange multipliers providing this conservation. Such form of the action with the Lagrange multiplier which providing current density conservation
\disn{r11}{
S=\int\! d^4 x\, \sqrt{-g}\ls j^\m \dd_\m\la - \sqrt{j^\m j^\n g_{\m\n}}\rs
\nom}
can be used in the description of an
perfect fluid with no vorticity
\cite{statja48}. The similarity between the action of embedding matter \eqref{r5} and
potentially moving perfect fluid
\eqref{r11} is not confined to the presence of Lagrange multipliers. Indeed, their structure is completely identical, including the presence of the root. The only difference is the presence of an additional index of the current density in \eqref{r5}, so  \eqref{r5} simply coincides with \eqref{r11} if $N=1$ since $\text{\bf tr}\sqrt{j_\m j^\al}=\sqrt{j_\m j^\m}$ if the vector $j^\m$ is timelike. Note that the theory which appearing when one uses  \eqref{r11} as $S^{\text{add}}$ in the action \eqref{v2a}  turns out to be equivalent to mimetic gravity.

The necessity of the consideration of continuity equation  $D_\m j^\m=0$ through addition of the term with Lagrange multiplier in the action of perfect fluid is a consequence of the continuous limit. If, on the contrary, one considers the matter as a set of individual particles which have definite worldlines, the continuity is present automatically. If one could construct an analogous "microscopic"{} description of the embedding matter which automatically provides the continuity equation \eqref{r6}, then it would be unnecessary to add the Lagrange term with $y^a$ to the action  \eqref{r5}. Unfortunately, it is still unclear how to do so.
The only thing that can be noted immediately is the possibility to completely exclude  $y^a$ (which plays the role of the Lagrange multiplier in this case) from the equations of motion by rewriting  \eqref{r7} in the equivalent form:
\disn{r12}{
D_\al\ls\hat\be_{\be\m}j^\m_a\rs-D_\be\ls\hat\be_{\al\m}j^\m_a\rs=0
\nom}
(the covariant derivatives here can be replaced by ordinary ones due to the symmetricity of connection). As a result the embedding function $y^a$ turns out to be completely excluded from the equations of motion of the embedding matter. Such matter then can be described by the conserving (in according to \eqref{r6}) currents $j^\m_a$ which also satisfy the equation of motion \eqref{r12} and contribute (see \eqref{r10} and \eqref{r3}) to Einstein equations
 \disn{r13}{
G^{\m\n}=\ka \ls T^{\m\n}+\be^{\m\n}\rs
\nom}
(we remind that $\be^{\m\n}$ and $\hat\be_{\m\n}$ are mutually inverse and can be expressed through $j^\m_a$ and metric).

An interesting observation is that the structure appearing in the action \eqref{r5} of embedding matter, namely the square root of the matrix product, is well known by the so-called bimetric gravity theories \cite{bimetric_review}. In these theories the role of matrices is played by two independent metrics, whereas in the action \eqref{r5}  one matrix is a metric $g_{\m\n}$ and another is a symmetric quadratic expression $j^\n_a j^{\al a}$. Among the different topics of study in bimetric theories the question of using non-standard root branches is considered along with other mathematical aspects connected with the presence of such singular expression in the theory, see, e.g. \cite{golovnev2017,hassan2018}. Such subtleties in the approach of embedding gravity require additional study.

By analogy with the possibility to write the perfect fluid action \eqref{r11} in many equivalent ways \cite{statja48}, there are several ways to write an action equivalent (possibly up to certain
special
solutions) to \eqref{r5} without the use of matrix root.
In this case, however, the number of independent variables increases. It reaches its maximum in the polynomial action
  \disn{r14}{
S_3^{\text{add}}=\int\! d^4 x\, \sqrt{-g}\,
\biggl( j^\m_a\dd_\m y^a-\be^{\m\n}g_{\m\n}+\frac{1}{2}\la_{\al\be}\ls\be^{\al\ga}g_{\ga\de}\be^{\de\be}-j^\al_a j^{\be a}\rs\biggr).
\nom}
Here the independent variables are $j^\m_a$, $y^a$ and symmetric tensors $\be^{\m\n}$, $\la_{\al\be}$. Variation w.r.t. them gives (besides the continuity equation\eqref{r6}):
 \disn{r15}{
\la_{\m\al}\be^{\al\be}g_{\be\n}+g_{\m\al}\be^{\al\be}\la_{\be\n}=2g_{\m\n},\qquad
\be^{\al\ga}g_{\ga\de}\be^{\de\be}=j^\al_a j^{\be a},\qquad
\dd_\m y^a=\la_{\m\n}j^{\n a}.
\nom}
The first of them has the solution $\la_{\m\n}=\be^{-1}_{\m\n}$, whose satisfaction leads to the fact that \eqref{r15} reproduces  \eqref{r7},\eqref{r8}.
It is also easy to check that in this case variation w.r.t. metric leads to Einstein equation \eqref{r13}. It should be noted that at the certain special values of $j^\m_a$ the solution of the first equation in \eqref{r15} is not unique, which leads to the special solutions which are absent in the standard formulation of the embedding theory. Such solutions possibly correspond to some solutions of  \eqref{r5} with non-standard choice of root branch.

The polynomial action \eqref{r14} is no more than quadratic w.r.t. $j^\m_a$ as well as $\be^{\m\n}$, moreover, there are no derivatives of these variables. By excluding certain quantities from \eqref{r14} one can obtain simpler (i.e. with less number of variables), but equivalent forms of action. If one varies the action \eqref{r14} w.r.t. $\be^{\m\n}$, obtaining the first equation of \eqref{r15}, then solve this equation for it as $\be^{\m\n}=\la^{-1}{}^{\m\n}$, again omitting the abovementioned special solutions, and substituting the result back in the action leads to
 \disn{r16}{
S_4^{\text{add}}=\int\! d^4 x\, \sqrt{-g}\,
\biggl( j^\m_a\dd_\m y^a-\frac{1}{2}\la^{-1}{}^{\m\n}g_{\m\n}-\frac{1}{2}\la_{\m\n}j^\m_a j^{\n a}\biggr).
\nom}
One can show that further exclusion $\la_{\m\n}$ from \eqref{r16} leads to \eqref{r5}. If one instead excludes $j^\m_a$ from \eqref{r16}, then  $j^\m_a=\la^{-1}{}^{\m\n}\dd_\n y_a$ and \eqref{r16} transforms to the above action \eqref{za1} where $\ta^{\m\n}$ is replaced by  $\la^{-1}{}^{\m\n}$.

Alternatively, the variable $j^\m_a$ can be excluded from \eqref{r14} in the first place. It gives the latter equation in \eqref{r15} and leads to another form of action:
 \disn{r17}{S_5^{\text{add}}=\int\! d^4 x\, \sqrt{-g}\,
\biggl(\frac{1}{2}\la^{-1}{}^{\m\n}(\dd_\m y^a)(\dd_\n y_a)+
\frac{1}{2}\la_{\al\be}\be^{\al\ga}g_{\ga\de}\be^{\de\be}-\be^{\m\n}g_{\m\n}\biggr).
\nom}
Further exclusion of $\la^{\m\n}$ from this action leads to the form of action which again contains matrix root:
 \disn{r18a}{
S_6^{\text{add}}=\int\! d^4 x\, \sqrt{-g}\,
\biggl(\text{\bf tr}\sqrt{(\dd_\si y^a)(\dd_\al y_a)\be^{\al\ga}g_{\ga\de}\be^{\de\be}}-\be^{\m\n}g_{\m\n}\biggr).
\nom}

\section{Conclusion and possible development}
Starting from the geometric description of the formulation of gravity proposed by Regge and Teitelboim as \emph{embedding theory}, one can rewrite the theory in the form of \emph{embedding gravity} \eqref{v2a},\eqref{r5} which is GR with some additional matter. Such a transition is analogous to one that can be done for \emph{mimetic gravity} \eqref{vi15},\eqref{v1}, when a theory, which was initially formulated as a result of certain change of variables in GR, is rewritten as GR with additional matter resembling a
perfect fluid with no vorticity
\eqref{v2},\eqref{p12}. Both approaches can be used in attempts to solve the dark matter and dark energy problems: for the mimetic gravity approach see \cite{mimetic-review17}; for the embedding theory approach this question was studied in the FRW approximation in the papers \cite{davids97,davids01,statja26}.

The approach of the embedding theory has a clear geometrical sense: spacetime is treated as a surface in the flat bulk, whereas the mimetic gravity is based on the change of variables \eqref{v1} which is constructed artificially to separate the conformal mode of the metric. In the original formulation of mimetic gravity the appearing fictional matter has very simple properties (it is a perfect fluid with potential motion), so the model must be modified (e.g. by introduction of additional parameters, see Introduction) to explain dark matter. On the contrary, the matter appearing in the embedding gravity approach, is highly nontrivial by itself, though has some properties in common with perfect fluid. Therefore an attempt to explain dark matter (and possibly dark energy) through embedding gravity with the above action seems promising.
Especially interesting problem is the construction of the abovementioned
(see before \eqref{r12})
"microscopic"{} description of the embedding matter which
could
automatically provide the satisfaction of continuity equation \eqref{r6} for a matter current.

One can use the mentioned similarity between perfect fluid and embedding matter to understand its properties more deeper. If the independent current density $j^\m_a$ is nonzero only at a single value of the index $a$ (e.g. $j^\m_a=j^\m\de^0_a$), then the embedding matter becomes perfect fluid precisely, see after \eqref{r11}. Therefore one can analyze the equations of embedding gravity
in the framework of "non-relativistic"{} approximation from the point of view of the ambient space (for each value of $\m$ ambient space vector $j^\m_a$ its 0th component prevails):
 \disn{r18}{
j^\m_a=j^\m\de^0_a+h^\m_a,
\nom}
where $h^\m_a$ is a small perturbation.
In zeroth order of $h^\m_a$ the equation \eqref{r12} (containing the inverse quantity $\hat\be_{\m\n}$ which needs to be regularized)
reduces to the geodesic equation for a normalized velocity vector of the main component of the embedding matter $u_\m=\hat\be_{\m\n}j^\m$, see \cite{statja48}. However, the perturbative analysis turns out to be nontrivial problem because of non-smooth dependence of $\be_\m{}^\al$ and $\hat\be_{\m\n}$ on $j^\m_a$ in the expansion \eqref{r18} (we remind that $\be_\m{}^\al$ is a result of matrix root taking in \eqref{r5} and $\hat\be_{\m\n}$ is an inverse of it). Such analysis requires additional study and lays beyond the scope of the present paper.

This nontriviality is largely related to the non-linearizability of Regge-Teitelboim equations \eqref{r2} (their properties were discussed, e.g. in \cite{statja33}) in the original formulation of the embedding theory when the most natural background embedding function $y^a(x)$ is chosen which corresponds to a 4D plane. Also note that because of this non-linearizability it is difficult to use the results obtained for the dynamics of the embedding matter in the framework of  FRW symmetry \cite{davids97,davids01,statja26} as a basis on the perturbation theory aimed to transcend this symmetry.

As a final remark we note that if one considers an action
 \disn{r19}{
\tilde S=S^{\text{b}}+S^{\text{add}},\qquad
S^{\text{b}}=-T\int\! d^4x\sqrt{-g}
\nom}
instead of \eqref{v2a} ($T$ is a brane tension) with any kind of $S^{\text{add}}$ discussed in the section~2, one obtains the description of 3-brane; and lowering the dimension from 4 to 2 leads to the Nambu-Goto bosonic string. To prove that one needs to write the equation of motion of the brane described by embedding function  $y^a(x)$ with the action $S^{\text{b}}[g_{\m\n}]$, where the metric is given by \eqref{r1}, in the form analogous to \eqref{r4}:
 \disn{r20}{
D_\m\Bigl(g^{\m\n}\dd_\n y^a\Bigr)=0.
\nom}
On the other side one must take into account that for all forms of $S^{\text{add}}$ considered in the section~2 the equations of motion provide the existence of \eqref{r1} for the metric and lead to satisfaction of \eqref{r4}, and the variation of $\tilde S$ w.r.t. metric gives the equations $T g^{\m\n}+\ta^{\m\n}=0$ instead of Einstein ones.

If \eqref{za1} is chosen as  $S^{\text{add}}$ and dimension is set to two, then the bosonic string action turns out to have the form
 \disn{r21}{
\tilde S=-T\int\! d^2x \ls\sqrt{-g}+\frac{1}{2}\Bigl( (\dd_\m y^a)(\dd_\n y_a) - g_{\m\n}\Bigr)\tilde\tau^{\m\n}\rs,
\nom}
where new independent variable $\tilde\tau^{\m\n}=-\sqrt{-g}\,\tau^{\m\n}/T$ is introduced. Interestingly enough that it is this action that leads to the Polyakov action \cite{polyakov1981} in a simplest way --- through exclusion of the variable $g_{\m\n}$. Indeed, the variation w.r.t. $g_{\m\n}$ gives the equation $\sqrt{-g}\,g^{\m\n}=\tilde\tau^{\m\n}$. In 2 dimensions it leads to the condition $\det\tilde\tau=-1$, and usage of the obtained equation in \eqref{r21} allows to rewrite the action in the following form:

 \disn{r22}{
\tilde S=-\frac{T}{2}\int\! d^2x\, \tilde\tau^{\m\n} (\dd_\m y^a)(\dd_\n y_a).
\nom}
Taking into account that an arbitrary $2\times2$ matrix satisfying $\det\tilde\tau=-1$ can be written as $\tilde\tau^{\m\n}=\sqrt{-h}\,h^{\m\n}$, where $h_{\m\n}$ is an arbitrary auxiliary metric, the action can be presented as
 \disn{r23}{
\tilde S=-\frac{T}{2}\int\! d^2x\, \sqrt{-h}\,h^{\m\n} (\dd_\m y^a)(\dd_\n y_a),
\nom}
i.e. in the Polyakov form.
Since a classical theory of bosonic string is much simpler than the GR, a theory with an action $\tilde S$ \eqref{r19} and different forms of $S^{\text{add}}$ from the section ~2  can be a toy model useful for the analysis of embedding gravity with the same  $S^{\text{add}}$ in the action \eqref{v2a}.

{\bf Acknowledgements.}
The authors are grateful to A.~Golovnev and A.~Starodubtsev for useful discussions.
The work of one of the authors (A.~A.~Sheykin) was supported by RFBR grant N~18-31-00169.


\end{document}